\documentclass{icrc2009}

\usepackage{graphicx}   
\usepackage[font=footnotesize]{subfig} 
\usepackage{fixltx2e}
\usepackage{url}

\newcommand{\shorttitle}[1]%
{\markboth{Proceedings of the 31\MakeLowercase{$^{st}$} ICRC, {\L}\'{o}d\'{z} 2009}{#1} }
\newcommand{\etal}{\MakeLowercase{\textit{et al. }}} 

\begin{document}
\title{Importance of high $p_t$ and nuclear physics for simulating UHECR Air Showers }

\author{\IEEEauthorblockN{Jeff Allen\IEEEauthorrefmark{1},
	Glennys R. Farrar\IEEEauthorrefmark{1}}
                            \\
\IEEEauthorblockA{\IEEEauthorrefmark{1}Center for Cosmology and Particle Physics, New York University, USA}
}

\shorttitle{Allen \etal }
\maketitle

\begin{abstract}
Observational evidence from Auger and earlier experiments shows a deficit of signal in a surface detector compared to predictions, which increases as a function of zenith angle, when the energy of the event is fixed by fluorescence measurements.  We explore three potential explanations for this: the ``Cronin effect" (growth of high-transverse momentum cross sections with nuclear size), the need for more particles at high transverse momentum in $p-p$ collisions than currently predicted by high energy hadronic models used for air shower simulations, and the possibility that secondary interactions in the target air nucleus produces additional relatively soft pions not included in simulations.  We report here on the differences between Pythia and QGSJet II, especially for high $p_t$ particles.  The possible impact of these effects on the predicted surface array signal are also reported.
\end{abstract}

\begin{IEEEkeywords}
simulation, airshower, muons
\end{IEEEkeywords}
 
\section{Introduction}
A discrepancy between observations and simulations of ultra-high energy air showers has been recognized for many years, initially evident as an energy-scale discrepancy between air-fluorescence and surface array experiments.  A discrepancy is also seen between the predicted and observed ``attenuation curve", plotting some measure of the surface array signal as a function of angle, for a constant intensity cut (e.g., for the N highest energy events at the given angle).  The discrepancy can be parameterized by a ``muon rescaling" factor $N_\mu$ and an energy rescaling factor, $E_{Resc}$ and the results of this are given using a number of different techniques in the presentation of A. Castellina for the Auger collaboration at ICRC 2009.  So far, efforts to remove this discrepancy by modifying the p-p event generators (QGSJet, EPOS, Sybill,...) have not succeeded to fit both the Auger observations and other data.  

Here, we investigate three different examples of ``ordinary physics" that are not included in present event generators, which can increase the signal in a surface array:  1)  the ``Cronin effect" \cite{cronin76}, 2) a systematic deficiency in the production of high $p_\perp$ particles in "minimum-bias" event generators like QGSJet compared to event generators like Pythia which have been tuned for high $p_\perp$ \cite{qgsjet, pythia}, and 3) the possible production of an excess of low-energy pions in the target-fragmentation region due to secondary scattering in the target nucleus.

 \begin{figure}[!t]
  \centering
  \includegraphics[width=2.0in]{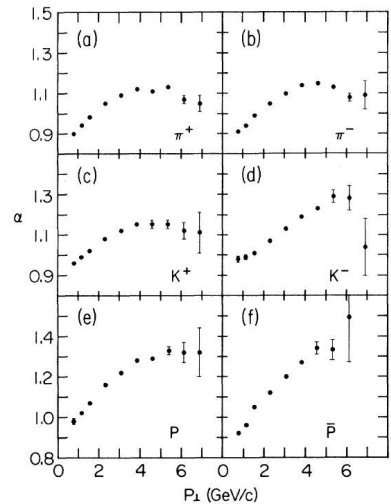}
  \caption{The ``Cronin effect" is the growth of $\alpha$, defined in the text, with $p_\perp$ such that it significantly exceeds $\alpha = 2/3$ \cite{cronin76} . }
  \label{alpha}
 \end{figure}

\begin{figure}[!t]
  \centering
  \includegraphics[width=2.5in]{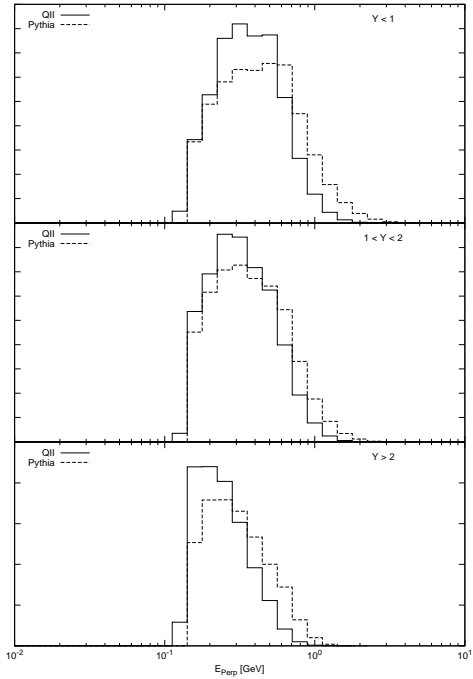}
  \caption{The predicted normalized distributions of $E_\perp$, for pions and kaons, of Pythia and QGSJet II, where $E_\perp$ is defined as $\sqrt{m^2 + p_\perp^2}$. The distributions are averaged over one million collisions of a $100 GeV$ proton upon a stationary proton. The top figure is the $E_\perp$ for secondary particles with a rapidity less than one. The middle figure is the $E_\perp$ distribution for secondary particles with rapidity between 1 and 2, and the bottom figure is the distribution for secondary particles with rapidity greater than 2.}
  \label{eperp}
 \end{figure}

\section{Cronin Effect}
\label{sec:cronin}
In the 1970's, it was discovered that nucleons are not just soft-mush but contain point-like constituents (now understood to be quarks) which are responsible for a power-law tail in high $p_\perp$ exclusive scattering, and in single particle inclusive production at high $p_\perp$.  As the physics of this phenomenon was explored experimentally and theoretically, one of the most puzzling discoveries was the ``Cronin effect" in which the cross-section for large-momentum transfer inclusive scattering was found to increase strongly with the atomic number of the target and the momentum transfer, compared to the behavior expected from the overall $A^{2/3}$ rescaling of the total cross section compared to $p-p$ scattering.  The dependence on $A$ and $p_\perp$ can be parameterized as
\begin{equation}
\label{cronin}
E \, \frac{d \sigma}{d^3\,p}_A =  A^{\alpha(p_\perp)} \, E \, \frac{d \sigma}{d^3\,p}_p. 
\end{equation}
Fig. \ref{alpha} shows $\alpha$ as a function of $p_\perp$.  The measurements were carried out at approximately $90^\circ$ in the CM, and done for beam energies of 200, 300 and 400 GeV and targets ranging up to Tungston.  Presumably, the Cronin effect results from multiple scattering, but it has never been satisfactorily explained theoretically and has not been further explored experimentally.  The dependence of $\alpha$ on the quark-content of the secondary is particularly fascinating.  \\

\section{Difference between QGSJet II and Pythia for $p_\perp \ge 0.5$ MeV}
\label{sec:PQ}
The various event generators have different predictions for the production of particles with large $p_\perp$. In particlular, Pythia, which is designed and tuned to describe high $p_\perp$ particles, predicts a larger signal at high $p_\perp$ than QGSJet II. The difference in the predictions of the two event generators are illustrated in Fig. \ref{eperp} for collisions of a $100 GeV$ primary proton on a stationary proton.  We see that QGSJetII does not produce enough high $p_\perp$ particles.

\begin{figure}[!t]
  \centering
  \includegraphics[width=2.0in, angle=270]{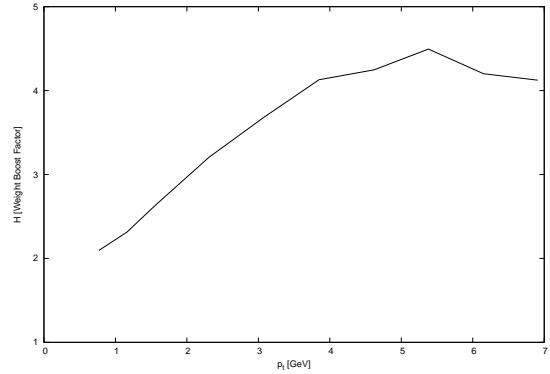}
  \caption{The reweighting factors, as a function of $p_\perp$, necessary to correct a target mass dependence of $\alpha=\frac{2}{3}$ to that of \cite{cronin76}, averaged over the secondary types predicted by QGSJetII.}
  \label{Hcronin}
 \end{figure}
 
\section{Method}
\begin{figure}[!t]
  \centering
  \includegraphics[width=2.5in]{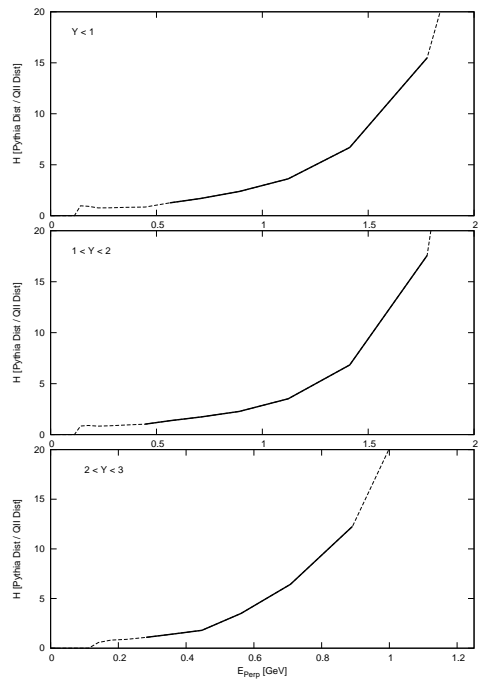}
  \caption{The ratio of the $E_\perp$ distributions of Pythia and QGSJet II, using the same interactions as Fig. \ref{eperp}, for three bins in rapidity. The solid line covers the region in which we modified the secondaries in QGSJet II interactions to match the $E_\perp$ distribution on Pythia.  }
  \label{Hpythia}
 \end{figure}

Since current simulations do not include the ``Cronin effect" and QGSJet does not produce enough high $E_\perp$ events even in p-p collisions, more realistic simulations would produce more secondaries at large scattering angle.  A muon originating from a typical production height of 5 km, with a $\approx 10^\circ$ production angle would arrive at 1 km.  To estimate the impact of increasing the number of high angle muons, we re-weight the secondaries in a suite of QGSJet simulations, by the factors shown in Figs. \ref{Hcronin} and \ref{Hpythia},  denoting these weight increases by $H_{Cronin}(p_\perp)$ and $H_{Pythia}(E_\perp)$ respectively.

 \begin{figure}[!t]
  \centering
  \includegraphics[width=3.5in, angle=270]{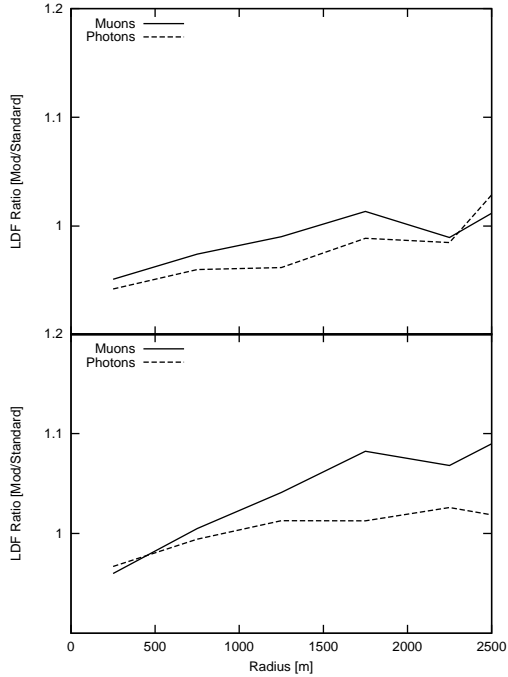}
  \caption{The ratio of the LDF with the Cronin-reweightings to the LDF when no change is made. The top figure is for a vertical shower, and the botton figure is for a shower inclined at $45^\circ$. For the muon ratios, what is plotted is the number density, while for photons what is plotted is the energy density.}
  \label{LDFcronin}
 \end{figure}

$H_{Cronin}(p_\perp)$ is found by taking the average of $14.4^{\alpha_i}$, where $i$ runs over particle species, weighted by the species abundance at high $p_\perp$. $H_{Pythia}(E_\perp)$ is found by taking the ratios the the $E_\perp$ distributions for Pythia and QGSJetII. The $E_\perp$ distributions were generated for incident energies of $100 GeV$, $1 TeV$, and $10 TeV$.

For every secondary particle produced in a QGSJetII interaction, the appropriate $H$ is applied to the individual weight of the secondary particle. In order to conserve energy, the weight of all secondaries produced in the collision is decreased by a common factor chosen so the final energy equals the initial energy.  This procedure effectively increases the number of high $p_\perp$ particles at the expense of multiplicity.

 \begin{figure}[!t]
  \centering
  \includegraphics[width=3.5in, angle=270]{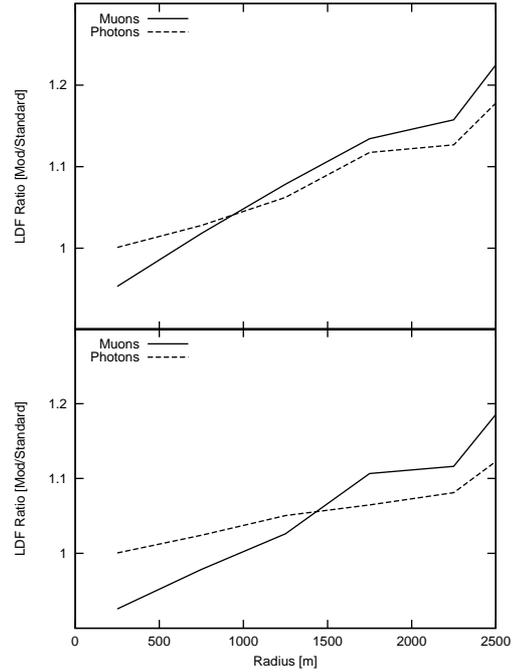}
  \caption{The ratio of the LDF with the Pythia-reweightings compared to the LDF from QGSJet. The top figure is for a vertical shower, and the botton figure is for a shower inclined at $45^\circ$. For the muon ratios, what is plotted is the number density, while for photons what is plotted is the energy density.}
  \label{LDFpythia}
 \end{figure}

\newpage

\newpage
\section{Results and Conclusions}
\begin{figure}[!t]
  \centering
  \includegraphics[width=2.0in, angle=270]{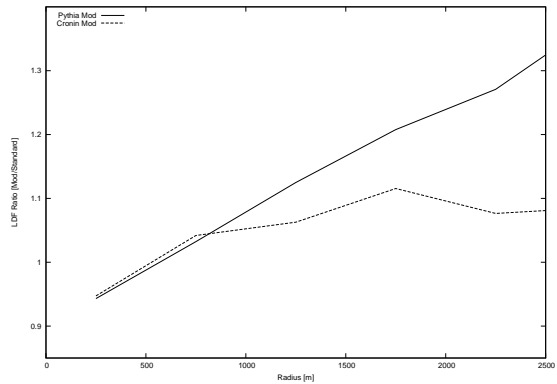}
  \caption{The ratio of the LDF energy density of muons for both the Pythia-reweightings and the implementation of the Cronin effect, for a shower inclined at $45^\circ$.}
  \label{ldfmueng}
 \end{figure}

\newpage
The impact on the Lateral Distribution Function, of the Cronin effect and of more accurately describing high pt particle production in the basic p-p collision, are shown in Figs. \ref{LDFcronin} and \ref{LDFpythia}.  Also shown in Fig. \ref{ldfmueng} are the ratio of the energy density of muons in the showers with and without reweighting.  Evidently, the effect on the energy at large core radius of extra high-pt pions and the muons from their decay can be strong, but the number of high $p_t$ secondaries is probably simply too small to explain the observed signal in the surface detector.  Thus, it seems that for the purposes of simulating extensive air showers, high-$p_t$ physics can safely be modeled crudely, without significant impact on the predicted surface array signal.  

Possibly more significant for the LDF is the production of relatively low energy pions by secondary particles from the primary collision which interact within the same nucleus before they escape.  The results of a study of this effect will be presented at ICRC.

\end{document}